\documentclass[preprint,english,aps,showpacs]{revtex4-1}
\pdfoutput=1
\usepackage{amsmath}
\usepackage{color}
\usepackage{graphicx}
\usepackage{epstopdf}

\usepackage{babel}

\begin{document}

\title{Facetted growth of Fe$_3$Si shells around GaAs nanowires on Si(111)}
\author{B.~Jenichen}
\email{bernd.jenichen@pdi-berlin.de}
\author{M.~Hilse}
\author{J.~Herfort}
\author{A.~Trampert}
\address{Paul-Drude-Institut f\"ur Festk\"orperelektronik,
Hausvogteiplatz 5--7,D-10117 Berlin, Germany}

\date{\today}

\begin{abstract}
GaAs nanowires and GaAs/Fe$_{3}$Si core/shell nanowire structures were grown by molecular-beam epitaxy on oxidized Si(111) substrates and characterized by transmission electron microscopy. The surfaces of the original GaAs NWs are completely covered by magnetic Fe$_3$Si, exhibiting nanofacets and an enhanced surface roughness compared to the bare GaAs NWs. Shell growth  at a substrate temperature of $T_{S}$~=~200~$^\circ$C leads to regular nanofacetted  Fe$_{3}$Si shells. These facets, which lead to thickness inhomogeneities in the shells, consist mainly of well pronounced Fe$_3$Si(111) planes. The crystallographic orientation of core and shell coincide, i.e. they are pseudomorphic. The nanofacetted  Fe$_3$Si shells found in the present work are probably the result of the Vollmer-Weber island growth mode of Fe$_3$Si on the \{110\} side facets of the GaAs NWs.
\end{abstract}

%%\pacs{68.70.+w, 68.55.ag, 68.37.Lp, 61.05.cp}
%%61.05.cp xrd, 68.37.Lp TEM, 68.55.ag semiconductors,68.70.+w, whiskers

%%\keywords{A1 Nanostructures; A3 Molecular beam epitaxy; B1 Gallium compounds; B1 Metals; B2 Magnetic materials}

\maketitle

\section{Introduction}

Nanowires that combine a semiconductor and a ferromagnet in a core/shell geometry have gained a lot of interest in recent years.\cite{Hilse2009,Rudolph2009,Rueffer2012,Dellas2010,Tivakorn2012,Yu2013} Because of the cylindrical shape of the ferromagnet, such core/shell nanowires could allow for a magnetization along the wire and thus perpendicular to the substrate surface. Ferromagnetic stripes or tubes with a magnetization perpendicular to the substrate surface have the potential for circular-polarized light emitting diodes that can optically transmit spin information in zero external magnetic field.\cite{Farshchi2011} This could enable three-dimensional magnetic recording with unsurpassed data storage capacities.\cite{Parkin2008,Ryu2012}
The combination of the binary Heusler alloy Fe$_{3}$Si (Curie temperature of about 840~K) and GaAs, has several advantages compared to most previously studied semiconductor/ferromagnet (SC/FM) core/shell NWs. The perfect lattice matching allows for the molecular beam epitaxy (MBE) growth of high quality planar hybrid structures.\cite{herfort03,Jenichen05,Herfort2006,Herfort2006b} In addition, the cubic Fe$_{3}$Si phase shows a robust stability against stoichiometric variations, which only slightly modify the magnetic properties.\cite{Herfort2004} Moreover, the thermal stability against chemical reactions at the SC/FM interface is considerably higher than that of conventional ferromagnets like Fe, Co, Ni, and Fe$_{x}$Co$_{1-x}$. \cite{herfort03}  We recently demonstrated that GaAs/Fe$_{3}$Si core/shell NWs prepared by MBE show ferromagnetic properties with a magnetization oriented along the NW axis (perpendicular to the substrate surface).\cite{hilse2013} However, the structural and magnetic properties of the core/shell NWs were found to depend strongly on the substrate temperature during the growth of the Fe$_{3}$Si shells,\cite{hilse2013,jenichen2014} and nanofacetting was observed.\cite{jenichen2014} In this work, we investigate periodically faceted Fe$_{3}$Si shells grown around GaAs(111) oriented cores and analyze the nanofacets by scanning electron microscopy (SEM), and transmission electron microscopy (TEM).

%%\cite{Hansen1958,Elliot1965}

\section{Experiment}

% fig.1
\begin{figure}[!t]
\includegraphics[width=16.0cm]{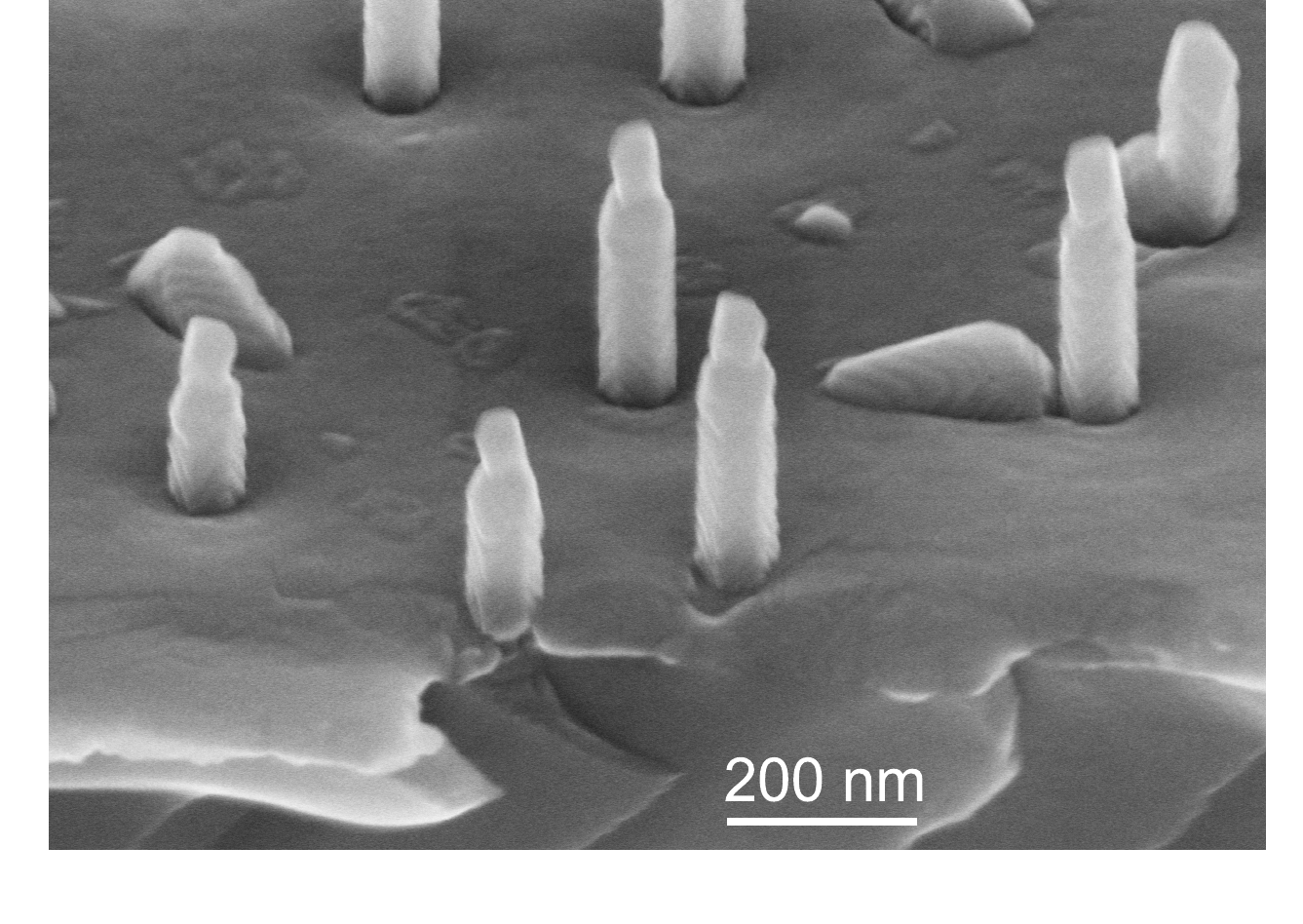}
\caption{SEM image of  GaAs/Fe$_{3}$Si core/shell  NWs and islands grown by molecular beam epitaxy on a Si(111) substrate.
}
\label{fig:SEM}
\end{figure}

GaAs/Fe$_{3}$Si core/shell NW structures were grown by MBE on Si(111) substrates. First, GaAs nanowires are fabricated by the Ga-assisted growth mode on  Si(111) substrates covered with a thin native Si-oxide layer. The growth mechanism is the vapor-liquid-solid (VLS) mechanism,\cite{Wagner1964,Mandl2006,glas2007,Fontcuberta2008,Wacaser2009} where pin holes in the SiO$_2$ serve as nucleation sites.\cite{Breuer2011}  A Ga droplet is the preferred site for deposition from the vapor. The GaAs NW then starts to grow by preferential nucleation at the spatially restricted GaAs/Si interface (IF). Further growth is nearly unidirectional and proceeds at the solid/liquid IF.  The GaAs NWs are grown at a substrate temperature of 580~$^\circ$C, with a V/III flux ratio of unity and an equivalent two-dimensional growth rate of 100~nm/h.
Once the GaAs NW templates are grown, the sample is transferred under ultra high vacuum conditions to an As free growth chamber for deposition of the ferromagnetic films.  There the GaAs NW templates were covered with Fe$_3$Si shells  at substrate temperatures varying between 100~$^\circ$C and 350~$^\circ$C.  The NW shells grown at  200~$^\circ$C show a regular nanofacet structure and were selected for a careful analysis of the nanofacets. More details regarding the growth conditions can be found in Ref.~\cite{hilse2013}.

The resulting core/shell NW structures were characterized by SEM and TEM. The TEM
specimens are prepared by mechanical lapping and polishing,
followed by argon ion milling according to standard techniques.
TEM images are acquired with a JEOL 3010 microscope operating at
300~kV. The cross-section TEM methods provide
high lateral and depth resolutions on the nanometer scale, however they average
over the thickness of the thin sample foil or the thickness of the NW as a whole.

\section{Results and Discussion}

% fig.2
\begin{figure}[!t]
\includegraphics[width=14.0cm]{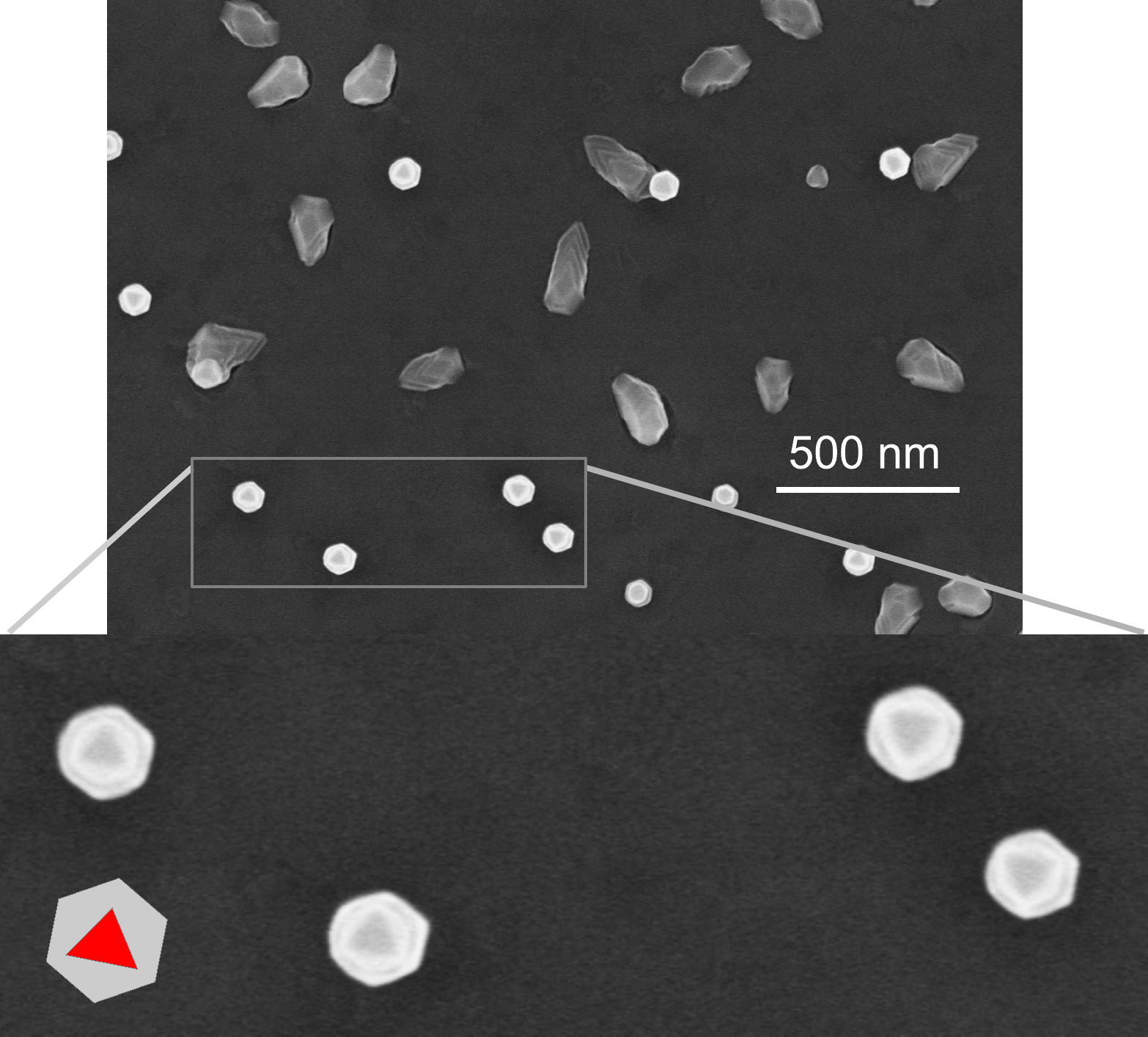}
\caption{(color online) SEM top view of  GaAs/Fe$_{3}$Si core/shell  NWs and islands between the NWs grown by molecular beam epitaxy on a Si(111) substrate.
}
\label{fig:SEM_top_view}
\end{figure}

Figure~\ref{fig:SEM} shows an SEM micrograph of the GaAs/Fe$_{3}$Si core/shell NWs.  A relatively low area density of well oriented NWs of $\sim$5$\times$10$^8~$cm$^{-2}$ is found as well as a comparable density of hillhocks.
During the last stage of GaAs NW growth no Ga is supplied, and so the remaining Ga in the droplet on top of the NWs is consumed, leading to an elongation of the NW at reduced diameter.\cite{jenichen2014} The sidewalls of the NWs exhibit nanofacetted surfaces.
Figure~\ref{fig:SEM_top_view} shows an SEM top view image of  GaAs/Fe$_{3}$Si core/shell  NWs grown by MBE on a Si(111) substrate. The NWs exhibit the typical hexagonal cross-section (sketched in gray),\cite{jenichen2014} however, at higher magnification we observe triangular features (sketched in red) which are connected to the thinner necks of the NWs, where the tilted Fe$_{3}$Si(111) planes form extended facets (cf. Fig.~\ref{fig:SEM})  intersecting the top Fe$_{3}$Si(111) plane, which is parallel to the substrate surface. This results in the formation of the triangular features. The lower edges of the necks are more rounded.

%$\langle$111\rangle_{Fe3Si}$
%3$\overline{3}$3
%108$^\circ$
%$\overline${2}$

% fig.3
\begin{figure}[!t]
\includegraphics[width=14.0cm]{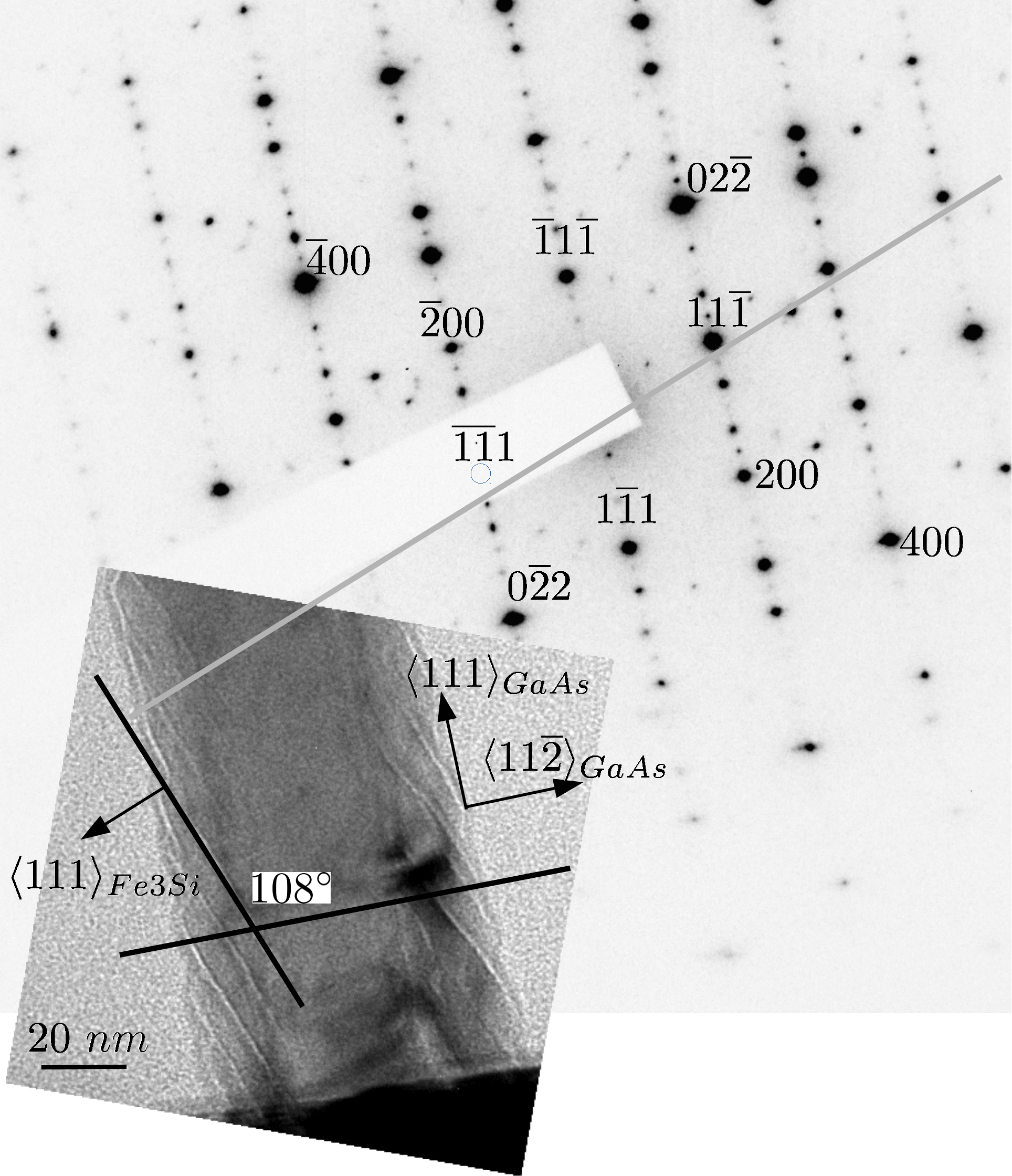}
\caption{Multi-beam bright-field TEM micrograph and the corresponding SAD pattern illustrating the orientational relationship of a GaAs/Fe$_{3}$Si core/shell NW. The straight line near the 11$\overline{1}$ reflection is oriented perpendicular to the nanofacets of Fe$_{3}$Si visible in the corresponding micrograph.
}
\label{fig:BF}
\end{figure}

Figure~\ref{fig:BF} shows a multi-beam bright-field TEM micrograph and corresponding SAD pattern, illustrating the orientational relationship of a GaAs/Fe$_{3}$Si core/shell NW on Si(111). We observe a coincidence of the core- and shell-orientations, hence the Fe$_{3}$Si growth is predominantly epitaxial on the GaAs. In addition, the crystallographic orientation of the Fe$_{3}$Si shell was confirmed by high-resolution TEM (not shown here). The grey line drawn on the SAD pattern near the 11$\overline{1}$ reflection is oriented perpendicular to the Fe$_{3}$Si nanofacets visible in the corresponding micrograph. The separation of the diffraction spots along this line corresponds to the (111) net plane distance of Fe$_{3}$Si  and indicates the nanofacets are mainly (111)-oriented. In the SAD pattern from a single core/shell NW the Fe$_{3}$Si maxima are stronger probably due to the larger volume fraction of the shell. In order to measure the SAD, the substrate was first oriented near the [011] zone axis. Then the sample had to be tilted a little further in order to reach the [011] zone axis of the GaAs NW, since the axis of the NW was not exactly perpendicular to the Si surface. In the NW SAD pattern shown in Fig.~\ref{fig:BF} the fundamental reflections of the Fe$_{3}$Si are more intense than the super-lattice maxima.\cite{Jenichen05} Nevertheless we can distinguish, 2$\overline{2}$2, 3$\overline{3}$3, and 4$\overline{4}$4 maxima indicating the NW is properly oriented and the crystallographic orientations of core and shell basically coincide. This illustrates that a growth temperature of 200~$^\circ$C results in a highly perfect Fe$_{3}$Si shell structure.
The surface nanofacets are inclined to the (111) net planes parallel to the Si surface by an angle of approximately 108$^\circ$. The formation of facets reduces the overall surface energy and evidences non negligible material transport over distances small compared to the NW lengths.\cite{wulff1901,dudley2010} Unfortunately, the orientation of those surface nanofacets does not coincide with  the GaAs NWs facets  corresponding to $\{110\}$ planes.\cite{grandal2014} As a result, the facetted growth leads to shell thickness inhomogeneity. On a larger length-scale the Fe$_{3}$Si shell is approximately reproducing the shape of the GaAs core NWs.\cite{jenichen2014}

Layer-by-layer growth could in principle solve the problem of nanofacetting. However, even planar Fe$_3$Si grows on GaAs(001) in the Vollmer-Weber (VW) island growth mode.\cite{kag09} Poor wetting during the growth of Fe$_3$Si on GaAs leads initially to isolated islands, despite of perfect lattices match.  We speculate that the nanofacetted  Fe$_3$Si shells found in the present work are a result of VW island growth. One way to improve the homogeneity and other structural properties of Fe$_{3}$Si-shells could be to use surfactants.
%%~\cite{Copel1989,LeGoues1990,hoegen1991,Tromp1992,vegt1992}
For planar growth on GaAs, surfactant materials like  Sb~\cite{Kageyama2004}, Bi~\cite{Tixier2003} and Te~\cite{Grandjean1992} have been investigated.

\section{Conclusions}

GaAs core NWs were grown epitaxially on the oxidized Si(111) surface (inside holes of the SiO$_2$ film) via the VLS growth mechanism. Then magnetic Fe$_3$Si shells were grown resulting in continuous covering of the cores.  Fe$_3$Si shells grown at a substrate temperature of $T_{S}$~= ~200~$^\circ$C are fully epitaxial with a nanofacetted surface. The ($\overline{1}$11) facets are most pronounced, forming a regular pattern around the GaAs NWs. This facetting is probably the result of the VW island growth mode of Fe$_3$Si on  GaAs.

\section{Acknowledgement}
The authors thank Claudia Herrmann for her support during the
MBE growth, Doreen Steffen for sample preparation, Astrid Pfeiffer
for help in the laboratory,  Anne-Kathrin Bluhm for the SEM micrographs, Ryan Lewis, Esperanza Luna and Uwe Jahn
for valuable support and helpful discussion.

\section{References}

%\bibliography{Zitate}
%merlin.mbs apsrev4-1.bst 2010-07-25 4.21a (PWD, AO, DPC) hacked
%Control: key (0)
%Control: author (8) initials jnrlst
%Control: editor formatted (1) identically to author
%Control: production of article title (-1) disabled
%Control: page (0) single
%Control: year (1) truncated
%Control: production of eprint (0) enabled
%

%%\newpage

\end{document}